\documentclass[aps,prl,twocolumn,showpacs,superscriptaddress]{revtex4}

\usepackage{graphicx}
\usepackage{dcolumn}
\usepackage{bm}
\usepackage{epsfig,}
\usepackage{multirow}

\newcommand{ \stopo}   {\mbox{$\tilde{t}_{1}$}}
\newcommand{ \stopone}    {\mbox{$\tilde{t}_{1}$}}
\newcommand{ \stoponeb}    {\mbox{$\bar{\tilde{t}}_{1}$}}
\newcommand{ \mstopo}   {\mbox{$m(\tilde{t}_1)$}}

\newcommand{\tauh}{\mbox{$\tau_{\rm h}$}}

\newcommand{\bbar}   {\mbox{$\overline{b}$}}

\newcommand{\ppbar}{\mbox{$p\overline{p}$}}

\newcommand{\ttbar}{\mbox{$t\overline{t}$}}

\newcommand{\met}{\mbox{${E\!\!\!\!/_{T}}$}}
\newcommand{\metvec}{\mbox{${\vec{E}\!\!\!\!/_{T}}$}}
\newcommand{\rp}{\mbox{${R_p}$}}

\newcommand{\rpv}{\mbox{${R\!\!\!\!\!\:/_p}$}}

\newcommand{\br}{\mbox{${\cal B}$}}
\newcommand{\intlum}{\mbox{${ \int {\cal L} \; dt}$}}
\newcommand{\pt}{\mbox{$p_{{T}}$}}

\newcommand{\et}{\mbox{$E_{{T}}$}}

\newcommand{\St}{\mbox{$S_{{T}}$}}
\newcommand{\abseta}{\mbox{$|\eta|$}}

\newcommand{\mt}{\mbox{$m_{{T}}$}}

\newcommand{\gev}  {\mbox{${\rm GeV}$}}
\newcommand{\gevc} {\mbox{${\rm GeV}/c$}}
\newcommand{\pgev} {\mbox{${\rm GeV}/c$}}
\newcommand{\gevcc}{\mbox{${\rm GeV}/c^2$}}

\newcommand{\invpb}{\mbox{${\rm pb}^{-1}$}}

\newcommand{\eps}{\mbox{$\epsilon$}}

\newcommand{ \Wjet}{\mbox{$W+{\rm jet}$}}
\newcommand{ \Njet}{\mbox{$N_{\rm jet}$}}

\newcommand{\GEANT}{{\sc geant}}
\newcommand{\PROSPINO}{{\sc prospino}}
\newcommand{\PYTHIA}{{\sc pythia}}

\def \etal     {\relax\ifmmode{et \; al.}\else{$et \; al.$}\fi}

\begin{document}



\title{Search for Pair Production of Scalar Top Quarks Decaying to 
a $\tau$ Lepton and a $b$ Quark in $p\bar{p}$ Collisions at $\sqrt{s}=1.96$ TeV}%



\affiliation{Institute of Physics, Academia Sinica, Taipei, Taiwan 11529, Republic of China} 
\affiliation{Argonne National Laboratory, Argonne, Illinois 60439} 
\affiliation{Institut de Fisica d'Altes Energies, Universitat Autonoma de Barcelona, E-08193, Bellaterra (Barcelona), Spain} 
\affiliation{Baylor University, Waco, Texas  76798} 
\affiliation{Istituto Nazionale di Fisica Nucleare, University of Bologna, I-40127 Bologna, Italy} 
\affiliation{Brandeis University, Waltham, Massachusetts 02254} 
\affiliation{University of California, Davis, Davis, California  95616} 
\affiliation{University of California, Los Angeles, Los Angeles, California  90024} 
\affiliation{University of California, San Diego, La Jolla, California  92093} 
\affiliation{University of California, Santa Barbara, Santa Barbara, California 93106} 
\affiliation{Instituto de Fisica de Cantabria, CSIC-University of Cantabria, 39005 Santander, Spain} 
\affiliation{Carnegie Mellon University, Pittsburgh, PA  15213} 
\affiliation{Enrico Fermi Institute, University of Chicago, Chicago, Illinois 60637} 
\affiliation{Comenius University, 842 48 Bratislava, Slovakia; Institute of Experimental Physics, 040 01 Kosice, Slovakia} 
\affiliation{Joint Institute for Nuclear Research, RU-141980 Dubna, Russia} 
\affiliation{Duke University, Durham, North Carolina  27708} 
\affiliation{Fermi National Accelerator Laboratory, Batavia, Illinois 60510} 
\affiliation{University of Florida, Gainesville, Florida  32611} 
\affiliation{Laboratori Nazionali di Frascati, Istituto Nazionale di Fisica Nucleare, I-00044 Frascati, Italy} 
\affiliation{University of Geneva, CH-1211 Geneva 4, Switzerland} 
\affiliation{Glasgow University, Glasgow G12 8QQ, United Kingdom} 
\affiliation{Harvard University, Cambridge, Massachusetts 02138} 
\affiliation{Division of High Energy Physics, Department of Physics, University of Helsinki and Helsinki Institute of Physics, FIN-00014, Helsinki, Finland} 
\affiliation{University of Illinois, Urbana, Illinois 61801} 
\affiliation{The Johns Hopkins University, Baltimore, Maryland 21218} 
\affiliation{Institut f\"{u}r Experimentelle Kernphysik, Universit\"{a}t Karlsruhe, 76128 Karlsruhe, Germany} 
\affiliation{Center for High Energy Physics: Kyungpook National University, Daegu 702-701, Korea; Seoul National University, Seoul 151-742, Korea; Sungkyunkwan University, Suwon 440-746, Korea; Korea Institute of Science and Technology Information, Daejeon, 305-806, Korea; Chonnam National University, Gwangju, 500-757, Korea} 
\affiliation{Ernest Orlando Lawrence Berkeley National Laboratory, Berkeley, California 94720} 
\affiliation{University of Liverpool, Liverpool L69 7ZE, United Kingdom} 
\affiliation{University College London, London WC1E 6BT, United Kingdom} 
\affiliation{Centro de Investigaciones Energeticas Medioambientales y Tecnologicas, E-28040 Madrid, Spain} 
\affiliation{Massachusetts Institute of Technology, Cambridge, Massachusetts  02139} 
\affiliation{Institute of Particle Physics: McGill University, Montr\'{e}al, Canada H3A~2T8; and University of Toronto, Toronto, Canada M5S~1A7} 
\affiliation{University of Michigan, Ann Arbor, Michigan 48109} 
\affiliation{Michigan State University, East Lansing, Michigan  48824} 
\affiliation{University of New Mexico, Albuquerque, New Mexico 87131} 
\affiliation{Northwestern University, Evanston, Illinois  60208} 
\affiliation{The Ohio State University, Columbus, Ohio  43210} 
\affiliation{Okayama University, Okayama 700-8530, Japan} 
\affiliation{Osaka City University, Osaka 588, Japan} 
\affiliation{University of Oxford, Oxford OX1 3RH, United Kingdom} 
\affiliation{University of Padova, Istituto Nazionale di Fisica Nucleare, Sezione di Padova-Trento, I-35131 Padova, Italy} 
\affiliation{LPNHE, Universite Pierre et Marie Curie/IN2P3-CNRS, UMR7585, Paris, F-75252 France} 
\affiliation{University of Pennsylvania, Philadelphia, Pennsylvania 19104} 
\affiliation{Istituto Nazionale di Fisica Nucleare Pisa, Universities of Pisa, Siena and Scuola Normale Superiore, I-56127 Pisa, Italy} 
\affiliation{University of Pittsburgh, Pittsburgh, Pennsylvania 15260} 
\affiliation{Purdue University, West Lafayette, Indiana 47907} 
\affiliation{University of Rochester, Rochester, New York 14627} 
\affiliation{The Rockefeller University, New York, New York 10021} 
\affiliation{Istituto Nazionale di Fisica Nucleare, Sezione di Roma 1, University of Rome ``La Sapienza," I-00185 Roma, Italy} 
\affiliation{Rutgers University, Piscataway, New Jersey 08855} 
\affiliation{Texas A\&M University, College Station, Texas 77843} 
\affiliation{Istituto Nazionale di Fisica Nucleare, University of Trieste/\ Udine, Italy} 
\affiliation{University of Tsukuba, Tsukuba, Ibaraki 305, Japan} 
\affiliation{Tufts University, Medford, Massachusetts 02155} 
\affiliation{Waseda University, Tokyo 169, Japan} 
\affiliation{Wayne State University, Detroit, Michigan  48201} 
\affiliation{University of Wisconsin, Madison, Wisconsin 53706} 
\affiliation{Yale University, New Haven, Connecticut 06520} 
\author{T.~Aaltonen}
\affiliation{Division of High Energy Physics, Department of Physics, University of Helsinki and Helsinki Institute of Physics, FIN-00014, Helsinki, Finland}
\author{J.~Adelman}
\affiliation{Enrico Fermi Institute, University of Chicago, Chicago, Illinois 60637}
\author{T.~Akimoto}
\affiliation{University of Tsukuba, Tsukuba, Ibaraki 305, Japan}
\author{M.G.~Albrow}
\affiliation{Fermi National Accelerator Laboratory, Batavia, Illinois 60510}
\author{B.~\'{A}lvarez~Gonz\'{a}lez}
\affiliation{Instituto de Fisica de Cantabria, CSIC-University of Cantabria, 39005 Santander, Spain}
\author{S.~Amerio}
\affiliation{University of Padova, Istituto Nazionale di Fisica Nucleare, Sezione di Padova-Trento, I-35131 Padova, Italy}
\author{D.~Amidei}
\affiliation{University of Michigan, Ann Arbor, Michigan 48109}
\author{A.~Anastassov}
\affiliation{Rutgers University, Piscataway, New Jersey 08855}
\author{A.~Annovi}
\affiliation{Laboratori Nazionali di Frascati, Istituto Nazionale di Fisica Nucleare, I-00044 Frascati, Italy}
\author{J.~Antos}
\affiliation{Comenius University, 842 48 Bratislava, Slovakia; Institute of Experimental Physics, 040 01 Kosice, Slovakia}
% ...
\collaboration{CDF Collaboration\footnote{With visitors from $^a$University of Athens, 15784 Athens, Greece, 
$^b$Chinese Academy of Sciences, Beijing 100864, China, 
$^c$University of Bristol, Bristol BS8 1TL, United Kingdom, 
$^d$University Libre de Bruxelles, B-1050 Brussels, Belgium, 
$^e$University of California Irvine, Irvine, CA  92697, 
$^f$University of California Santa Cruz, Santa Cruz, CA  95064, 
$^g$Cornell University, Ithaca, NY  14853, 
$^h$University of Cyprus, Nicosia CY-1678, Cyprus, 
$^i$University College Dublin, Dublin 4, Ireland, 
$^j$University of Edinburgh, Edinburgh EH9 3JZ, United Kingdom, 
$^k$University of Heidelberg, D-69120 Heidelberg, Germany, 
$^l$Universidad Iberoamericana, Mexico D.F., Mexico, 
$^m$University of Manchester, Manchester M13 9PL, England, 
$^n$Nagasaki Institute of Applied Science, Nagasaki, Japan, 
$^o$University de Oviedo, E-33007 Oviedo, Spain, 
$^p$Queen Mary, University of London, London, E1 4NS, England, 
$^q$Texas Tech University, Lubbock, TX  79409, 
$^r$IFIC(CSIC-Universitat de Valencia), 46071 Valencia, Spain, 
}}
\noaffiliation



\begin{abstract}
We search for pair production of supersymmetric top quarks ($\tilde{t}_{1}$),
followed by {\it R}-parity violating decay $\tilde{t}_{1}\rightarrow\tau b$ with a branching ratio $\beta$, using 322 pb$^{-1}$ of $p\overline{p}$ collisions 
at $\sqrt{s}=1.96$ TeV collected by the CDF II detector at Fermilab. 
Two candidate events pass our final selection criteria, consistent with the standard model expectation.
We set upper limits on the cross section 
$\sigma(\tilde{t}_{1}\bar{\tilde{t}}_{1})\times\beta^2$ as a function of the stop mass $m(\tilde{t}_1)$.
Assuming $\beta=1$, we set a 95\% confidence level limit $m(\tilde{t}_1) > 153$ GeV$/c^2$. 
The limits are also applicable to the case of a third generation scalar leptoquark ($LQ_3$) decaying $LQ_3 \to \tau b$.

\end{abstract}


\pacs{12.60.Jv, 14.80.Ly,13.85.Rm,11.30.Pb,11.30.Er}  

\maketitle


In supersymmetric (SUSY) models~\cite{mssm}, the spin-1/2 quarks and leptons have 
spin-0 quark and lepton partners. 
Experimental data suggest that the superpartners of the first and second generation are heavier than those of the standard model (SM) particles,
while the mass of the lighter scalar top quark (stop or \stopone) is weakly constrained and can be below that of 
the top quark~\cite{lstop}.
This is due to the mixing between the left and right handed interaction eigenstates 
which is a function of the large Yukawa coupling of the top quark. 
At the Fermilab Tevatron stop quarks and antiquarks can be produced in pairs in strong interactions 
($gg/q\bar{q} \to \stopone\stoponeb$). A single stop could also be produced at the Tevatron, e.g., via $b g \to \stopone \tau$~\cite{singlestop}; however, unlike pair production, this process requires an $R$-parity (\rp) violating vertex. 
In regions of parameter space not excluded by data, \rp\ violating (\rpv) couplings are small~\cite{rpvtheory}, making single stop production negligible compared to pair production. 
Stop quarks can decay into lighter SUSY and SM particles if \rp\ is conserved or into ordinary quarks and/or leptons if \rp\ is violated. 
Within the framework  of \rpv\ SUSY~\cite{rpvtheory}, theoretical studies indicate that 
the dominant decay mode for the light stop is the lepton number violating decay $\stopone \to \tau b$
for a wide range of SUSY model parameters, 
including the region allowed by neutrino oscillation data~\cite{rpvstopbr}.

Leptoquarks appear in various SM extensions~\cite{lqtheory}.
Charge $|Q|=2/3$ and $|Q|=4/3$ third generation scalar leptoquarks ($LQ_3$) are expected to decay into $\tau$ and $b$ with $\br(LQ_3 \to \tau b)=1$ for all $LQ_3$ states when $m(LQ_3)<m(t)$.
The next-to-leading order (NLO) cross section for $LQ_3\overline{LQ}_3$ production is very close to the \stopone\stoponeb\ production cross section $\sigma(\stopone\stoponeb)$
since diagrams with virtual gluino exchange are strongly suppressed with the existing limits on gluino mass~\cite{gluinomass}.
Thus, the limits obtained for \rpv\ stop should be fully applicable to the $LQ_3$ case.

In this Letter we describe a search for $\stopone\stoponeb \to \tau^{+}\tau^{-} b\bbar$ with the upgraded CDF detector (CDF II)~\cite{cdf2det} and set an upper limit on  $\sigma(\stopone\stoponeb)\times\beta^2$, neglecting additional decay modes that may pass selections of this analysis when $\beta\equiv\br(\stopone \to \tau b)<1$. 
We look for a final state with either an electron or muon from the decay $\tau\rightarrow\ell\nu_{\ell}\nu_{\tau}$ ($\ell=~e$ or $\mu$), 
a hadronically decaying tau $\tauh$, missing transverse energy \met~\cite{cdfdefs} from the neutrinos, and two or more jets.  
We have studied the addition of a requirement that the jets are consistent with originating from the hadronization of a $b$ quark but found that the increase in purity is outweighed by the loss in signal acceptance. 
Therefore, we make no such specific requirement.
This analysis uses approximately three times more data at a higher $\sqrt s$ than the previous CDF result~\cite{cdfRPVstop122} 
that set a 95\% C.L. limit of $\mstopo>122$ \gevcc.
The increased $\sqrt{s}$ is expected to give a substantial increase in the $\stopone\stoponeb$ production rate, e.g., $\sim 35\%$ for $\mstopo=155$ \gevcc.


CDF II features several main subsystems critical to this analysis.
The charged particle tracking system consists of multi-layer silicon detectors and an open-cell cylindrical drift chamber enclosed in a $1.4$ T superconducting magnet.
At $|\eta| < 1$~\cite{cdfdefs} charged particle trajectories traverse all chamber layers, while at larger $|\eta|$ the chamber coverage is reduced progressively.
The calorimeter system is organized into electromagnetic and hadronic sections segmented in a projective tower geometry 
and covers $\abseta < 3.6$. A set of strip and wire chambers is located within the central electromagnetic calorimeter at approximately the depth of shower maximum and aids in reconstructing electrons, and photons for $\abseta < 1.1$.
The muon detection system is located outside of the calorimeter and covers $\abseta < 1.0$.


The analysis begins with a data sample collected by inclusive lepton-plus-track triggers~\cite{cdf2LT}.
These triggers select events containing an electron (muon) candidate with $\et > 8\ \gev$ ($\pt > 8\ \pgev$) and
a second track, which is required to be consistent with originating from a tau decay 
by demanding that there be no other nearby tracks with $\pt > 1.5\ \pgev$ 
between the cones of $0.175$ and $0.524$ radians around the track. 
The integrated luminosity of the data sample is $322\pm19$~\invpb~\cite{cdf_lumi}.

From the trigger sample we select events offline by identifying at least one lepton with
$\pt^{\ell} >10\ \gevc$ 
and at least one $\tauh$ candidate in $|\eta|<1$. 
The details of the $\tauh$ identification algorithm can be found in Refs.~\cite{ztt-prd,mythesis}.
We require $p_T^{\tau}> 15$ \gevc.
Jets are reconstructed using a fixed-cone algorithm with  $\Delta R \equiv \sqrt{ \Delta \eta^2 + \Delta \phi^2 } = 0.4$ within $\abseta < 2.4$. 

The dominant SM backgrounds are vector boson production, QCD, and $t\bar{t}$ production. 
In QCD multijet events, for example, semileptonic $b$ quark decays or $\gamma$ conversions can be misidentified 
as lepton candidates, and narrow jets can be misidentified as \tauh\ candidates.
We require the sum of the $\pt $ of the tracks within $\Delta R <0.4$ around the lepton candidate ($I_{trk}$) be less than 2 GeV/$c$, 
and no jet with $\et > 15\ \gev$ within $0.3< \Delta R <0.8$ around the lepton.
Further, we reject events where the muon or electron candidate 
is consistent with a cosmic ray muon or $\gamma$ conversion electron (see Ref.~\cite{ztt-prd} for details).
We veto events where the invariant mass of the primary electron (muon) and a reconstructed electron (muon) candidate, 
which is required to pass loose identification criteria~\cite{ztt-prd},
is $76 < m_{\ell\ell} < 106\ \gevcc$.
We also reject events with $76 < m_{e\tau} < 106\ \gevcc$ and azimuthal separation of $|\Delta \phi_{e\tau}| > 2.9$ rad.
For the muon channel we do not apply a similar requirement, as the probability for a
muon to be reconstructed as a $\tauh$ is negligible.
To suppress further QCD and $Z\to\tau\tau$ events~\cite{cdfRPVstop122}, we require
$\St  \equiv |{\vec{p}_T}^{\ \ell}| + |\vec{p}_T^{\ \tauh}| + |\vec{\met}|/c  > 110\ \gevc$.

We define six regions in the $\mt(\ell,\met) \equiv \sqrt{2p_T^{\ell}E\!\!\!\!/_{T}(1-\cos\Delta\phi_{\ell,E\!\!\!\!/_{T}})}$ versus \Njet\ plane, 
and denote them as A$_j$ (B$_j$) for $\mt\leq 35$ \gevcc\ ($\mt>35$ \gevcc) and $j=0$, $1$ or $2$ for $\Njet=$~0, 1 or $\geq$ 2.
We count into \Njet\  the jet candidates that have $\et > 20\ \gev$ and are separated from any of $e$, $\mu$ or $\tauh$ by $\Delta R > 0.8$.
The minimal values of $\St $ and jet $\et $ are optimized for maximum significance 
in the A$_2$ region for $ 140 - 160$ \gevcc\ \stopone's.
The  $\mt \leq 35\ \gevcc$ cut effectively separates signal from \Wjet\ and $\ttbar$ backgrounds.
The $\Njet\geq 2$ requirement strongly suppresses the Drell-Yan contribution.
The data in region A$_2$ were not examined until the analysis procedure was finalized.
The regions with $\Njet=$~0 or 1 contain mostly background and were used mainly as control samples for validation.
Region B$_2$ has an appreciable signal acceptance ($\sim 40\%$ of that in region A$_2$) but substantially higher background expectation.
For statistical interpretation of the data, we developed a likelihood method that, in addition 
to our primary signal region A$_2$, utilizes side-band regions A$_0$, B$_0$, and B$_2$, 
which are used to perform data-driven \Wjet\ background estimations and to improve the sensitivity of the analysis.

The total event acceptance is
$\alpha \equiv \eps_{\rm MC} \cdot \eps_{\rm trig}$.
Here $\eps_{\rm MC}$ is the product of geometrical and kinematical acceptances, efficiencies to identify lepton and \tauh\ candidates, including isolation requirements, and the efficiency for the all remaining cuts.
We use \PYTHIA\ Monte Carlo (MC) generator~\cite{pythia} and the \GEANT3-based \cite{geant} CDF II detector simulation to calculate $\eps_{\rm MC}$.
Our nominal choice for the parton distribution functions (PDFs) is {\sc CTEQ5L}~\cite{cteq} 
with the renormalization scale $Q \equiv \sqrt{m(\stopone)^2 +\pt(\stopo)^2}$.
The trigger efficiency $\eps_{\rm trig}$ is measured using data~\cite{mythesis}.
In region A$_2$ $\alpha$ increases nearly linearly 
from about $0.6$\% at $\mstopo=100$ \gevcc\ to $2.7$\% at 170 \gevcc\ and 
is similar for both electron and muon channels.


The combined systematic uncertainty on $\alpha$ decreases almost linearly from 11\% for \mstopo = 100 \gevcc\ to 7.2\% for 170 \gevcc\ 
and is similar in both channels.
The largest contribution comes from the PDF systematic uncertainty, which is estimated using the uncertainty sets
of {\sc CTEQ6.1M} PDFs~\cite{cteq} and the technique described in Ref.~\cite{pdf_uncert}.
For a $150\ \gevcc$ stop this uncertainty on $\alpha$ is 4.0\%.
The uncertainty due to an imperfect knowledge of the jet energy scale, 
determined by varying the scale by $\pm 1\sigma$, is 2.9\%. 
The uncertainty due to the amount of initial and final state radiation
is found to be 2.5\%. 
Other sources of systematic uncertainty include the uncertainties in lepton and $\tauh$ identification and isolation, 
and \met\ resolution, and amount to a 5.1\% relative contribution.
The uncertainty on the integrated luminosity  is 6\%.


The SM backgrounds come from two sources: (i)~events with a true $\ell\tauh$ pair
from $Z/\gamma^*(\to \tau\tau$)+jets,
$\ttbar$ and diboson ($WW$, $WZ$, $ZZ$) production;
and (ii)~events where lepton or $\tauh$ candidates do not originate from a true lepton or tau but
from the jets in \Wjet, $Z/\gamma^*(\to \ell\ell$)+jets and QCD multijet events. 
We first estimate the background from SM processes excluding the \Wjet\ contribution.
Drell-Yan, $\ttbar$, and $WW$ production 
are estimated using \PYTHIA\ and the CDF II detector simulation.
For Drell-Yan backgrounds we use scale factors that improve the agreement between the prediction for the yield of these events in MC simulation and the yield observed in data.
The QCD multijet contribution is estimated by extrapolating the number of observed events in data for events with non-isolated leptons, defined by $2<I_{trk}<10\ \gevc$, into the class of events with an isolated lepton, defined by $I_{trk}<2\ \gevc$~\cite{ztt-prd}.
The NLO cross sections of $6.7\pm0.7$~pb~\cite{ttbar-nlo} and $13.5\pm0.5$~pb~\cite{diboson-nlo} for \ttbar\ and $WW$ production, respectively, are used.
The contributions from $WZ$ and $ZZ$ are found to be negligible.

The \PYTHIA\ MC simulation does not accurately predict the absolute
rate of the \Wjet\ background contribution ($N^{W}$) in this analysis. 
Therefore, to estimate $N^{W}$ in each region,
we use the differences between the data and all other backgrounds plus signal in regions 
A$_2$, B$_2$, A$_0$, and B$_0$ and the assumption that 
${\cal R} \equiv \left [ N^{W}({\rm A}_2)/N^{W}({\rm B}_2) \right] \cdot \left[N^{W}({\rm B}_0)/N^{W}({\rm A}_0) \right] \sim 1$.
The ratios in ${\cal R}$ are determined by kinematics of the \Wjet\ events at fixed \Njet\ and are well modeled in MC. 
Based on MC predictions and cross checks with data vs MC comparisons, we conclude that ${\cal R}=1.0 \pm 0.5$ is a conservative assumption.
We define a likelihood function using Poisson statistics as 
a function of $\sigma(\stopone\stoponeb)$ and $N^{W}$.
The input parameters to the likelihood are the numbers of observed and expected events in each of the four regions.
The number of expected events in region $i$ is given by 
$N_i$ = $\sigma(\stopone\stoponeb)\cdot \br(\tau\tau \to \ell\tauh) \cdot \intlum \cdot \alpha_i$ + $N^{\rm BG}_i$ + $N_i^{W}$,
where the branching ratio $\br(\tau\tau \to \ell\tauh)\simeq 0.23$, 
$N^{\rm BG}_i$ includes all SM backgrounds except \Wjet\ events, and
$\alpha_i$ is the total event acceptance for signal in region $i$.
The ratio ${\cal R}=1.0 \pm 0.5$ and $N^{W}$ in regions A$_0$, B$_0$, and B$_2$ are nuisance parameters with flat prior distributions.
The large uncertainty on ${\cal R}$ does not affect our final results because $N^{W}($A$_2)$ is expected to be small.
We use this two-dimensional likelihood to estimate $N^{W}$ for each region 
and to calculate upper limits on $\sigma(\stopone\stoponeb)\times\beta^2$.

\begin{figure}[t]
\begin{center}
\centerline{
\epsfig{figure=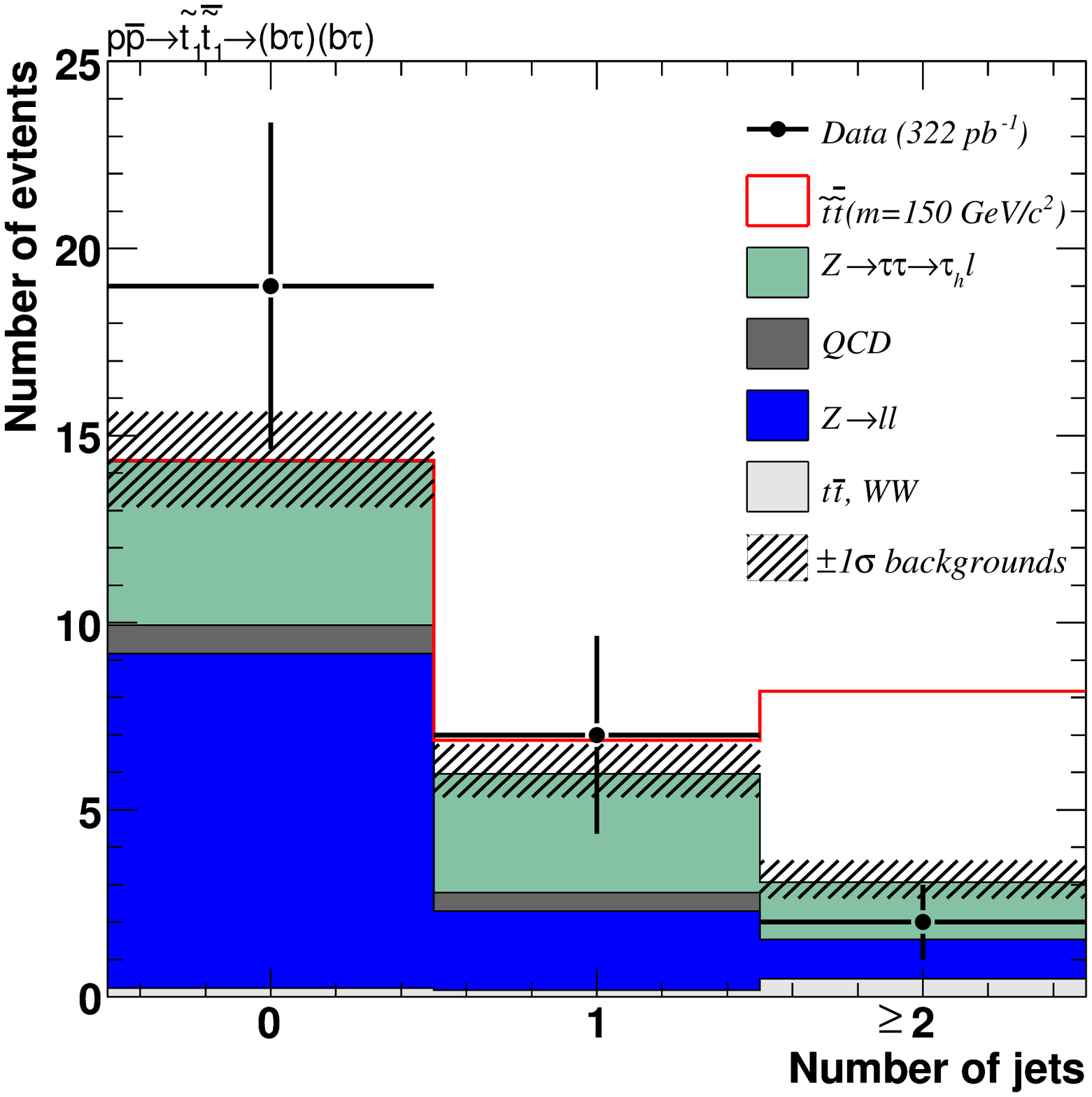,width=2.7in}
}
\caption{
Distribution of the number of jets ($\et > 20\ \gev$)
for events with $\mt(\ell,\met) \leq 35\ \gevcc$ (regions A$_0$, A$_1$, and A$_2$) 
compared to the expectations from SM processes and prediction for $\stopone\stoponeb$ ($\mstopo = 150\ \gevcc$) signal.
\label{njet}
}
\end{center}
\end{figure}

In Table~\ref{4region-data} we show the number of events observed in the data along with the SM expectation. 
In Fig.~\ref{njet} we present the \Njet\ distribution for events with $\mt \leq 35\ \gevcc$ (regions A$_0$, A$_1$, and A$_2$).

\begin{table}[t]
\caption{
Number of events observed in data, $N_{\rm obs}$, along with the expected number of SM background events.
The \Wjet\ contributions are shown separately as they are estimated using the observed number of events in the data, 
the SM prediction excluding the \Wjet\ contribution, and a possible stop quark contribution.
\label{4region-data}}
\begin{center}
\begin{tabular}{ c | c | c  c | c | c c }
\hline
\hline
   & \multicolumn{3}{c|}{$e+\tauh$ Channel} & \multicolumn{3}{c}{$\mu+\tauh$ Channel} \\       
Reg    &  $N_{\rm obs}$ & \multicolumn{2}{c|}{SM Backgrounds} & $N_{\rm obs}$ & \multicolumn{2}{c}{SM Backgrounds} \\       
       &   &  Other & $W+$jet       &   &  Other  & $W+$jet \\
\hline
A$_2$ & {\bf 1} &  $2.0^{+0.5}_{-0.4}$ & $0   ^{+0.4}_{-0  }$  & {\bf 1} & $1.0^{+0.4}_{-0.2}$ & $0   ^{+0.5}_{-0  }$  \\
B$_2$ &  4      &  $2.8^{+0.5}_{-0.3}$ & $1.0 ^{+2.0}_{-1.0}$  &  4      & $2.3^{+0.4}_{-0.3}$ & $1.7 ^{+2.0}_{-1.5}$  \\
A$_1$ &  4      &  $3.3^{+0.5}_{-0.5}$ & $0.2 ^{+1.2}_{-0.2}$  &  3      & $2.6^{+0.6}_{-0.4}$ & $0.1 ^{+0.8}_{-0.1}$  \\
B$_1$ &  9      &  $2.3^{+0.4}_{-0.3}$ & $6.7 ^{+3.2}_{-2.7}$  &  6      & $2.3^{+0.5}_{-0.3}$ & $3.8 ^{+2.7}_{-2.1}$  \\
A$_0$ & 11      &  $9.1^{+1.2}_{-1.1}$ & $1.6 ^{+2.7}_{-1.6}$  &  8      & $5.2^{+0.7}_{-0.5}$ & $2.5 ^{+2.4}_{-2.1}$  \\
B$_0$ & 25      &  $4.5^{+0.7}_{-0.6}$ & $21.1^{+5.6}_{-4.3}$  & 28      & $5.4^{+0.8}_{-0.6}$ & $23.6^{+4.9}_{-5.7}$  \\
\hline
\hline
\end{tabular} 
\end{center}
\end{table}

Two events are found in region A$_2$, consistent with the SM prediction.
We use the likelihood function 
to obtain a probability of such an observation given a specific signal cross section,
and calculate a 95\% C.L. limit on $\sigma(\stopone\stoponeb)\times\beta^2$.
The electron and muon channels are treated as two separate measurements,
taking into account the correlations among the systematic uncertainties. 

\begin{figure}[t]
\begin{center}
\centerline{
\epsfig{figure=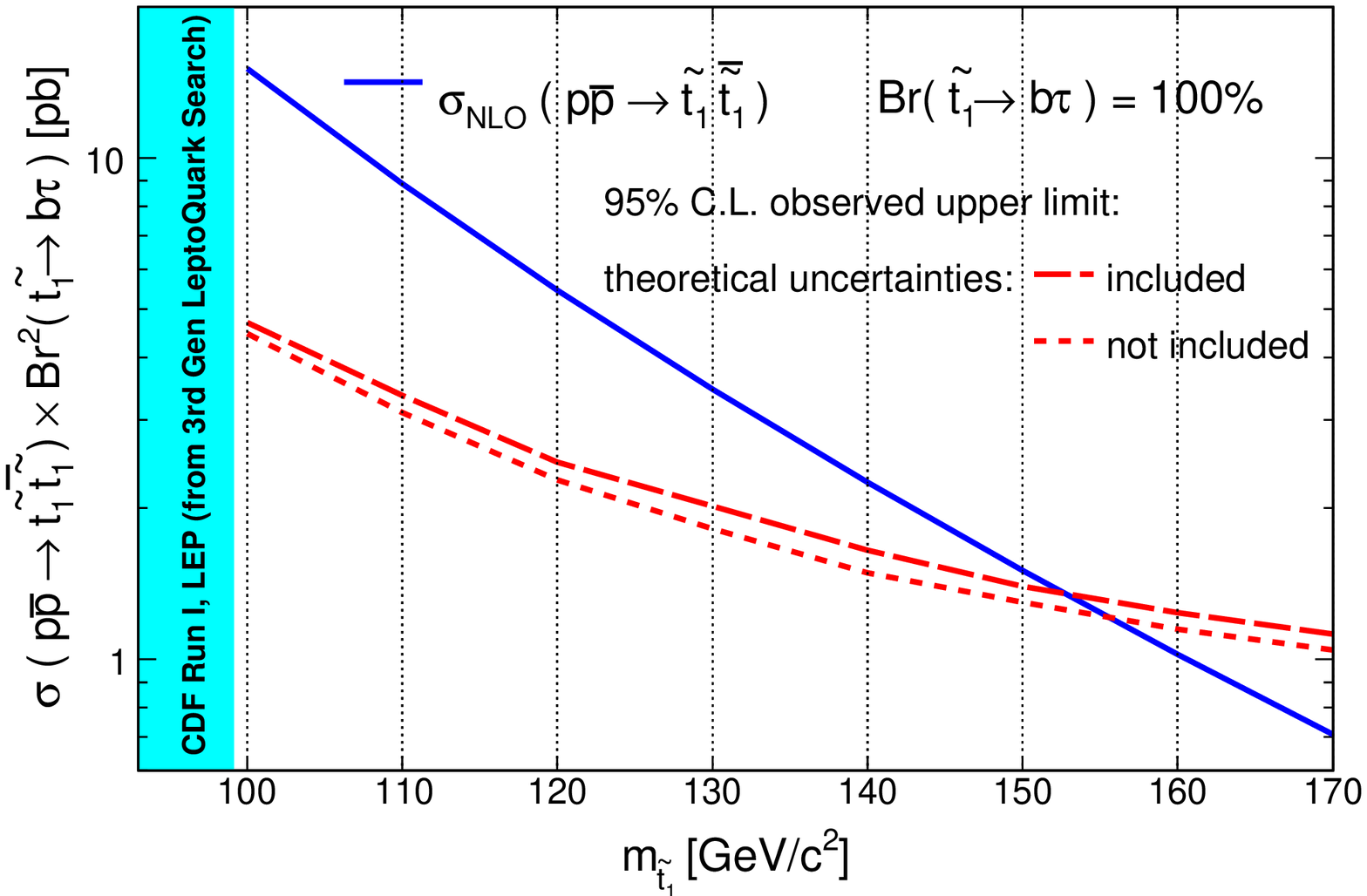,width=3.1in}}
\caption{95\% C.L. limit curves for the $\stopone \stoponeb$ production cross section times the branching ratio $\stopone \rightarrow b \tau$ squared
for the cases when the theoretical uncertainty on the cross section is (dashed line) and is not (dotted line) considered in the limit calculation (see text for details).
The previous constraint $m(LQ_3) > 99\ \gevcc$~\cite{lq3cdf1lep} is also shown.
\label{limit-example-plot}
}
\end{center}
\end{figure}

Figure~\ref{limit-example-plot} shows the 95\% C.L. limit curves for $\sigma(\stopone\stoponeb)\times\beta^2$ as a function of \mstopo, 
with the numerical results given in Table~\ref{limit_final}.
The dotted curve is our experimental result, compared to the NLO cross section (solid line) obtained 
using \PROSPINO\ version 2~\cite{prospino} with our nominal choice of {\sc CTEQ6.1M} PDFs~\cite{cteq}, and $Q$.
The theoretical uncertainty of $\pm 18\%$ on $\sigma(\stopone\stoponeb)$ is due to the choice 
of $Q$ (varying the scale from its nominal value by a factor of two or a half) and PDFs. 
Taking this uncertainty into consideration, the limits are re-evaluated and are shown in Fig.~\ref{limit-example-plot} using a dashed line.
The corresponding mass limits for the first and second cases are $156$ \gevcc\ (compared to 122 \gevcc~\cite{cdfRPVstop122}) and 153 \gevcc\ respectively.

\begin{table}[t]
\caption{
95\% C.L. upper limit on $\sigma(\stopone\stoponeb)\times\beta^2$ (in pb) as a function of $\mstopo$ for the cases when uncertainty on the theoretical cross section is considered ($\sigma^{95\%}_{with\ uncert}\times\beta^2$) and is not considered ($\sigma^{95\%}_{no\ uncert}\times\beta^2$),
where $\beta\equiv\br(\stopone \to \tau b)$.
\label{limit_final}
}
\begin{center}
\hspace*{-0.2cm}
\begin{tabular}{ lllllllll }
\hline
\hline
\mstopo\  (\gevcc) & 100 & 110 & 120 & 130 & 140 & 150 & 160 & 170 \\ 
\hline
$\sigma^{95\%}_{with\ uncert}\times\beta^2$ (pb)  & 4.73 & 3.37 & 2.50 & 1.99 & 1.61 & 1.38 & 1.26 & 1.14 \\
$\sigma^{95\%}_{no\ uncert}\times\beta^2$ (pb)    & 4.48 & 3.11 & 2.27 & 1.81 & 1.47 & 1.26 & 1.16 & 1.04 \\
\hline
\hline
\end{tabular}
\end{center}
\end{table}


In conclusion, we have searched for $\stopone\stoponeb$ production in the final state of
a lepton ($e$ or $\mu$), a hadronically decaying tau, and two jets using 322~\invpb\ of $\ppbar$ collision data
at $\sqrt{s}$ = 1.96 TeV.
We observed no excess of events in data over the number of expected events from SM processes.
In an \rpv\ SUSY scenario, we set a 95\% C.L. lower limit on the $\stopone$ mass of 153 \gevcc\ taking into account the theoretical uncertainties on the NLO cross section and assuming $\br(\stopone \to \tau b)=1$.
These results are also applicable to $LQ_3$ pair production.


We thank the Fermilab staff and the technical staffs of the participating institutions for their vital contributions. This work was supported by the U.S. Department of Energy and National Science Foundation; the Italian Istituto Nazionale di Fisica Nucleare; the Ministry of Education, Culture, Sports, Science and Technology of Japan; the Natural Sciences and Engineering Research Council of Canada; the National Science Council of the Republic of China; the Swiss National Science Foundation; the A.P. Sloan Foundation; the Bundesministerium f\"ur Bildung und Forschung, Germany; the Korean Science and Engineering Foundation and the Korean Research Foundation; the Science and Technology Facilities Council and the Royal Society, UK; the Institut National de Physique Nucleaire et Physique des Particules/CNRS; the Russian Foundation for Basic Research; the Comisi\'on Interministerial de Ciencia y Tecnolog\'{\i}a, Spain; the European Community's Human Potential Programme; the Slovak R\&D Agency; and the Academy of Finland.



\begin{thebibliography}{99}
\bibitem{cdfdefs}
 We use a coordinate system where $\theta$ and $\phi$ are the polar
  and azimuthal angles, respectively, 
  with respect to the proton beam direction ($z$ axis). 
  The pseudorapidity $\eta$ is defined as  $-\ln [ \tan(\theta/2) ]$.  
  The transverse momentum of a particle is denoted as
  $\pt = p \sin\theta$.  
  Similarly, the transverse energy is defined as $\et = E \sin\theta$. 
  The uncorrected missing transverse energy, $\met^{unc}$, is the magnitude of
  $\metvec^{unc} \equiv  -\sum E_{T}^i \,\hat{n}_i$, 
  where $\hat{n}_i$ is the unit vector in the
  transverse plane pointing from the interaction point to
  the energy deposition in calorimeter cell~$i$.
  \met\ is $\met^{unc}$ further corrected for the muon $\pt$ and for the effects 
  of non-ideal response of the calorimeter to $e$, $\tau$, and jets.
\end{thebibliography}
\end{document}